\documentclass[preprint, floatfix]{revtex4}

\usepackage{graphicx}     
\usepackage{dcolumn}      
\usepackage{bm}           

\begin{document}

\title{Catalyst-free selective area growth of InN nanocolumns by MBE}

\author{C. Denker}
\author{J. Malindretos}
\email{malindretos@ph4.physik.uni-goettingen.de}
\author{B. Landgraf}
\author{A. Rizzi}
\affiliation{IV. Physikalisches Institut, Georg-August-Universit\"at G\"ottingen, Germany}

\date{\today}

\begin{abstract}
Selective area growth of InN nanorods by plasma assisted molecular beam epitaxy is demonstrated.
Molybdenum is found to be a suitable mask material at the low substrate temperature of 475$\,^\circ$C needed for the growth of InN nanocolumns. The growth of arrays of single nanorods on a Si(111) substrate has been achieved with a thin molybdenum mask lithographically patterned with holes smaller than 60~nm. 
\end{abstract}



\maketitle

Nanocolumns have been studied intensively over the last years. Part of their attraction comes from the notion that they could be used as bottom-up building blocks for future opto- and nanoelectronics \cite{thelander2006nbo, lu2007nfb}. Another aspect which makes nanocolums very appealing for technological applications and basic research, is the fact that combinations of lattice mismatched materials can be realized with high crystal quality. This is due to the large free surface of the nanocolumns, which can efficiently relieve strain and thus relaxes the lattice matching constraint in epitaxial growth. In particular InN and its ternary alloys with Ga and Al can be grown with much more flexibility in the form of nanocolumns as compared to conventional layers or heterostructures. This material system is highly interesting for optoelectronic and light harvesting applications, since the fundamental bandgap can be tuned from the telecommunication wavelengths in the near infrared through all the visible spectrum up to the the near ultra-violett range. The self-organized formation of InN nanocolumns in molecular beam epitaxy (MBE) was observed by several groups \cite{grandal2007ami, stoica2006pai, wang2007sse, shen2006nip, denker2008sog}. These structures were grown on Si(111), on which an amorphous interlayer of SiN$_\textnormal{x}$ forms at the beginning of the growth, or on crystalline surfaces of lattice mismatched substrates like GaN or Si(111) with an AlN buffer layer. In all cases, the nanocolumns are reported to possess a high crystal quality, but self-organized growth provides only very little control over the size, the position and the shape of the nanostructures. This might also induce a spatial variation of alloy composition, which leads for instance to broad multicolor emission in InGaN/GaN nanocolumn LEDs  \cite{kishino2007ing}. Selective area growth (SAG) would allow for the growth of single, well separated columns with a controlled size, position and shape.

Reports on SAG of III-nitrides in MBE are still rare \cite{geelhaar2007aar, kishino2008itm}. Selective area growth of InN in MBE was only shown on a hole-patterned GaN template by Nanishi and coworkers \cite{harui2008tem, araki2009fpc}, who used a focused ion beam to etch pitches into the substrate. On Si(111) substrates, SAG has been demonstrated for MBE grown GaN nanocolums by Kishino \textit{et al.} \cite{kishino2008sag, kishino2008itm}. In that work the authors used a titanium mask, but high substrate temperatures of at least 900$\,^\circ$C are necessary for the occurrence of SAG. However, the much higher N$_2$ equilibrium pressure for InN requires much lower growth temperatures of $T_S \approx 475\,^\circ$C. At higher temperatures InN starts to decompose and significant desorption sets in at around 600$\,^\circ$C \cite{dimakis2005hgi}.

\begin{figure}[b]
	\centering
		\includegraphics[width=6cm]{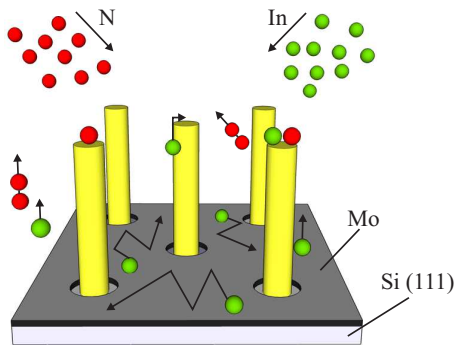}
	\caption{Principle mechanism of selective area growth by masking. The impinging atoms on the mask diffuse to the nucleation centers forming in the openings. Indium desorption is expected to play a minor role at the low growth temperature of InN.}
	\label{fig:SAG}
\end{figure}

In this letter, we report on the selective area growth of InN nanorods on n-Si(111). SAG was obtained by the use of patterned metallic masks (Fig.~\ref{fig:SAG}). Molybdenum and titanium were used as mask materials and were deposited with an electron beam evaporator under high-vacuum conditions. Depending on the size of the structure, optical or e-beam lithography was used to define the pattern in the mask. The thickness of the deposited material was 10~nm and 50~nm for e-beam and optical lithography, respectively. A lift-off step was performed after the deposition process. Since the metal films were exposed to air, a thin oxide layer forms at the surface. After mounting the samples into a Veeco GenII MBE system, they were heated up to 600$\,^\circ$C. It has been revealed by in-situ x-ray photoemission spectroscopy (not shown here) that the native oxide is changed, but not completely removed during this procedure. A standard effusion cell was used for indium and a Veeco UNI-BULB plasma source supplied the activated nitrogen. For all samples the same N-rich growth conditions were applied: a nitrogen flux of $F_{\mathrm{N}}=1.5$~sccm, a nitrogen plasma excitation power of $P_{\mathrm{N}}=450$~W, an indium beam equivalent pressure of $p_{\mathrm{In}}=3\cdot 10^{-8}$~mbar, a substrate temperature of $T_S=475\,^\circ$C (as measured by a thermocouple) and a growth duration of $t=1$~h. The growth was started by opening both shutters simultaneously within 2~min after striking the plasma. The growth was ended by closing both shutters and the substrate was cooled down to room-temperature at a rate of $-20\,^\circ$C/min. The morphology of the samples was analyzed using a high-resolution scanning electron microscope (SEM).

\begin{figure}
	\centering
		\includegraphics[width=8cm]{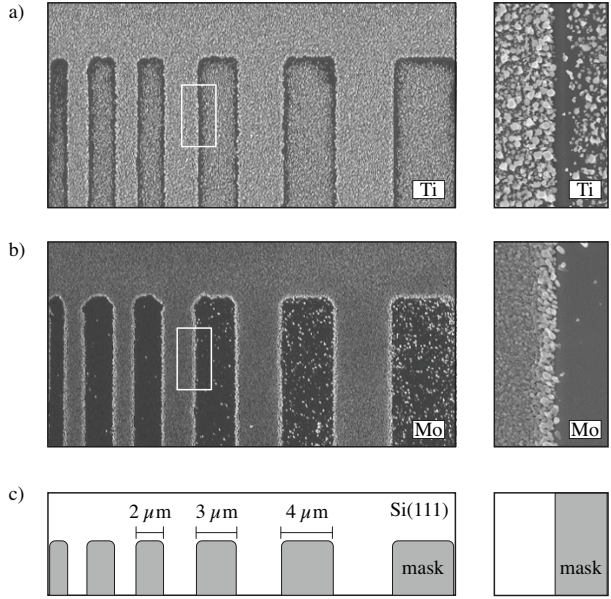}
	\caption{SEM images of titanium (a) and molybdenum (b) stripes on Si (111) after the growth of InN. Large areas free of nucleation are found on the molybdenum mask. The geometry of the mask is sketched in (c).}
	\label{fig:Fig-Micro}
\end{figure}

\begin{figure}
	\centering
		\includegraphics{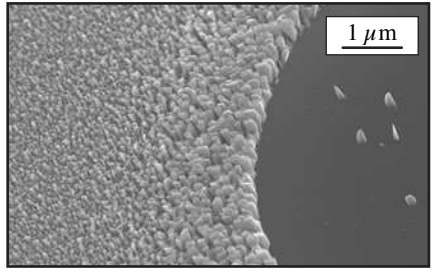}
	\caption{SEM image of the transition from the Si(111) substrate with InN nanorods (left) to a circular molybdenum mask (right). An increased amount of InN material is deposited along the border.}
	\label{fig:Fig-Trans}
\end{figure}

Figure~\ref{fig:Fig-Micro} shows the morphology of the molybdenum and titanium patterned Si (111) surfaces after growth. The mask patterns consist of stripes with different widths in the range of 1~$\mu$m to 4~$\mu$m. On the titanium mask, only a modest selectivity is visible. There is a region of about 100~nm free of nucleation along the edge of the mask. This can be seen in the magnified view of the titanium mask edge (Fig.~\ref{fig:Fig-Micro}a, right). However, the areas without any nucleation are very small. The molybdenum mask on the other hand induces a much higher selectivity. On stripes with a width of up to 2~$\mu$m only very few nuclei are observed. Large areas free of any nucleation can be found. The magnified view of the molybdenum mask edge (Fig.~\ref{fig:Fig-Micro}b, right) reveals that the amount of InN material deposited on the Si(111) substrate increases towards the border to the mask. The atoms diffusing from the mask to the substrate induce a gradient in the amount of InN material. This can also be seen in Fig.~\ref{fig:Fig-Trans}. With increasing width of the mask stripes, more and more material nucleates on the mask as well and the density of nanorods on the molybdenum increases (Fig.~\ref{fig:Fig-Micro}). Based on these observations, it can be concluded that the selective area growth is based on diffusion in the present case and that the length of diffusion on the molybdenum mask is at least one order of magnitude larger than the corresponding one on titanium. Therefore the most effective selectivity can be obtained with the molybdenum mask.

\begin{figure}[htb]
	\centering
		\includegraphics{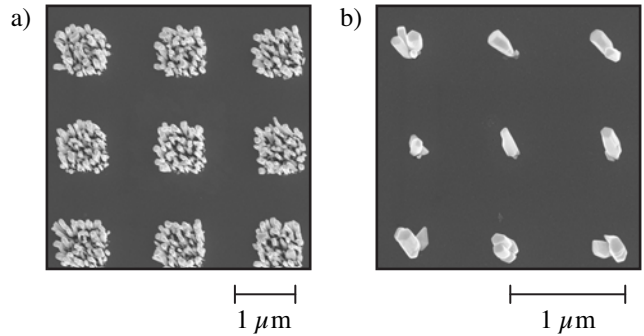}
	\caption{Selectiv area growth on a molybdenum mask with different hole diameters. On openings of 600~nm (a) arrays of InN nanorods are obtained while smaller openings of 30-90~nm (b) facilitate the growth of single InN nanorods.}
	\label{fig:Fig-openings}
\end{figure}

Next, arrays of InN nanorods were grown on a nano-patterned molybdenum mask with different hole sizes ranging from 600~nm down to less than 60~nm in diameter. Successful selective area growth is demonstrated without nucleation on the mask material. By changing the diameter of the holes the number of nanorods in the array can be controlled. In this way it is even possible to achive the growth of single, well separated InN nanorods for hole diameters of less than 60~nm (Fig.~\ref{fig:Fig-openings}).

In some cases even the smallest holes prepared contain more than one rod (Fig.~\ref{fig:Fig-openings}b). This is most probably due to slight variations in the actual diameter of the mask openings, since the lithography and the deposition process have not been fully optimized, yet. Furthermore, one can see that the nanorods do not exhibit a perfect alignment of their $c$-axes with the substrate. This tilting is well known from the self-organized growth of InN nanorods on Si(111) terminated with an amorphous toplayer \cite{denker2008sog}. The introduction of an AlN-buffer layer might result in a better alignment of the rods \cite{grandal2007ami} and is presently under investigation.

In conclusion, molybdenum is a suitable mask material for the selective area growth of InN nanorod arrays by MBE at a low substrate temperature of 475~$^\circ$C. Nucleation free areas of about $2 \times 2$~$\mathrm{\mu m^{2}}$ can be obtained. Furthermore, position controlled growth of single nanorods is demonstrated using a molybdenum mask with a hole sizes of less than 60~nm. The mechanism of the selective area growth is mainly based on diffusion and is still under investigation.

\begin{acknowledgments}
This work was supported by the ERANET project ÒNanoSci-ERA: NanoScience in the European Research AreaÓ of the EU FP6. We thank Michael Carsten for the XPS measurements.
\end{acknowledgments}

\end{document}